\documentclass[twocolumn,showpacs,amssymb,floatfix]{revtex4}
\usepackage{graphicx}
\usepackage{dcolumn}
\usepackage{bm}

\begin{document}

\title{Spin rotation for ballistic electron transmission
induced by spin-orbit interaction}
\author{Evgeny N.~Bulgakov $^1$ and  Almas. F.~Sadreev,$^{1,2}$}
\affiliation{$^1$ Institute of Physics, Academy of Sciences, 660036 Krasnoyarsk,
Russia,
$^2$ Department of Physics and Measurement Technology,
Link\"{o}ping University, S-581 83 Link\"{o}ping, Sweden}
\begin{abstract}
We study spin dependent electron transmission through one- and
two-dimensional curved waveguides and quantum dots with account of
spin-orbit interaction. We prove that for a transmission through
arbitrary structure there is no spin polarization provided that
electron transmits in isolated energy subband and only two leads
are attached to the structure.  In particular
there is no spin polarization in the one-dimensional wire for
which spin dependent solution is found analytically. The solution
demonstrates spin evolution as dependent on a length of wire.
Numerical solution for transmission of electrons through the
two-dimensional curved waveguides coincides with the solution for
the one-dimensional wire if the energy of electron is within the
first energy subband. In the vicinity of edges of the energy
subbands there are sharp anomalies of spin flipping.
\end{abstract}
\maketitle

\section{Introduction}
The electron spin precession phenomena at zero magnetic field
induced by a variable spin-orbit interaction (SOI) in 2DEG systems
was firstly proposed by Datta and Das \cite{datta} as a way for
the realization of the spin transistor. For this, the spin
precession is controlled via the Razhba SOI associated with the
interface electric field present in the GaAs heterostructures
that contains the 2DEG channel \cite{razhba}
\begin{equation}
\label{soi} V_{SO}^{\alpha} \, = \,\hbar\alpha [\hat{p}_x
\sigma_{y}-\hat{p}_y\sigma_{x}].
\end{equation}
The reason of spin precession is that the spin operators do not
commutate  with the SOI operator what leads to the spin evolution
for electron transport. In particular the SOI has a polarization
effect on particle scattering processes \cite{davydov} and this
effect was considered for different geometries of confinement of
the 2DEG
\cite{voskoboynikov,bulgakov,moroz,pichugin,molenkamp,mireles}.

The most simple case of the stripe geometry with the x-axis along
the stripe and the z-axis perpendicular to the stripe gives the
following transformation of spin state after transmission
\begin{equation}
\label{polarization}
\left(\matrix{1\cr
0\cr}\right)\Rightarrow\left(\matrix{\cos\theta/2\cr
\sin\theta/2\cr}\right)
\end{equation}
where \cite{datta,mireles}
\begin{equation}
\label{theta} \theta=2m^{*}\alpha L
\end{equation}
and $L$ is a length of the stripe. Therefore, the Razhba SOI
induces a spin precession of the transmitted electrons. Notice
that the spin precession is energy independent. This result is
valid if the confinement energy $\hbar^2/2m^{*}d^2$, where $d$ is
a width of the stripe, is much larger than the spin-splitting
energy induced by the SOI, and therefore, the intersubband mixing
is negligible \cite{mireles}. For the strong SOI the spin rotation
angle $\theta$ becomes to be the Fermi energy depend for ballistic
transport of electrons in the quasi one-dimensional wires and
stripes \cite{moroz,mireles}. Whatever the Razhba SOI leads to the
spin precession in the $(x, z)$ plane. Here we consider similar
phenomena for electron transmission through the curved waveguide
and quantum dots. The main difference between the straight
waveguide and curved one is that the spin rotation is given by two
angles.

Next, we find out conditions under which there is no a spin
polarization of transmitted electrons. We imply that a flow of
incident electrons have no spin polarization. By the spin
polarization we consider the mean spin\\ $<\sigma_{\alpha}>,
\alpha = x, y, z $ averaged over the electron flow. In particular
for the transmission through quantum dot we show a principal role
of the third lead for the spin polarization.

\section{The spin-orbit interaction in the inhomogeneous
two-dimensional case}

We write the total Hamiltonian of a confined 2DEG as
\begin{equation}
\label{total} H =
-\frac{\hbar^2}{2m^{*}}\left(\frac{\partial}{\partial x^2} +
\frac{\partial}{\partial y^2}\right)+V(x,y)+V_{SO}
\end{equation}
where $V(x,y)$ is the lateral confining potential.  Following to
Moroz and Barnes \cite{moroz} we assume that the SOI operator
$V_{SO}$ is formed by three contributions
$$
V_{SO} = V_{SO}^{\alpha} + V_{SO}^{\gamma} +
V_{SO}^{\alpha\alpha}.
$$
The first $V_{SO}^{\alpha}$ is related
to the Razhba SOI (\ref{soi}), in which the SOI constant $\alpha$
proportional to the macroscopic interface-induced electric field
is considered as constant. The second contribution $V_{SO}^{\gamma}$
to the SOI
comes from the electric field ${\bf E}(x,y)$ related to the
confining potential.

In order to derive the second contribution to the SOI we begin
with general description of SOI \cite{drell}
\begin{equation}
\label{SOIgen} V_{SO} = -\frac{e}{4m^2c^2}\left\{ \sigma({\bf E}
\times\hat{{\bf p}})+ \frac{i\hbar}{2}\sigma(\nabla\times {\bf
E})\right\}.
\end{equation}
For microscopic electric field ${\bf E}$ the second term in
(\ref{SOIgen}) equals zero. However for model cases of the
confining potential $V(x,y)$ the electric field
can violate an equality $\nabla\times{\bf E}=0$. In this case
the second term in (\ref{SOIgen}) is necessary to provide a
hermiticity of the total SOI operator.

For a 2DEG confined at semiconductor heterostructure interface we
can reduce the z-coordinate performing average over electron wave
function $\psi_0(z)$ strongly localized along the z-direction
\begin{equation}
\label{integral}
 V_{SO}\Rightarrow \int dz \psi_0(z) V_{SO}\psi_0(z).
\end{equation}
 As a result we obtain
$$ V_{SO}^{\gamma}=-\gamma\Bigg\{\sigma_z(E_x\hat{p}_y-E_y
\hat{p}_x) - E_z(\sigma_x\hat{p}_y-\sigma_y\hat{p}_x)$$
\begin{equation}
\label{soi2dd}
- \frac{i}{2}\hbar\sigma_z\left(\frac{\partial{E_x}}{\partial y}-
\frac{\partial{E_y}}{\partial x}\right)
-\frac{i}{2}\hbar\left(\sigma_y\frac{\partial{E_z}}{\partial
x}-\sigma_x \frac{\partial{E_z}}{\partial y}\right)\Bigg\}.
\end{equation}
Here electric field components are considering in a meaning of
integral (\ref{integral}) and depend on $x, y$ only.

For particular case of straight wire directed along the y-axis
with the lateral confining potential $U = U(x)$ we obtain from
(\ref{soi2dd}) the expression given by Moroz and Barnes (formula
(5) in \cite{moroz}).They used a parabolic approximation for the
confining potential. Here we consider a popular hard wall
approximation and imply the following confining potential $$
U(x)=\cases{0,\quad if \quad |x| < d/2, \cr U_0, \, if \quad |x|
\geq d/2}.$$
Then substituting the electric field $E_x = -U'(x)$
into (\ref{soi2dd}) we have
\begin{equation}
\label{UU} V_{SO}^{\gamma}(x)= \hbar k\gamma\sigma_z sign(x) U_0
\delta(x\mp d/2).
\end{equation}
For $|x| > d/2$ we have from the Schr\"odinger equation the
following solution
\begin{equation}
\label{psix}
\psi(x)=C\exp\left(-\frac{\sqrt{2m^{*}(U_0-E)}}{\hbar}x\right)
\end{equation}
where $C$ is the normalization constant. Using a property of delta
function that a difference between derivatives of the wave
function at the right and left of the delta function is obeying to
$\Delta \psi'(\pm d/2))=\pm 2m^{*}k\sigma_z\gamma U_0 \psi(\pm d/2)$ we have
from (\ref{psix}) that $$\Delta \psi'(\pm d/2)) \rightarrow 0$$
for $U_0\rightarrow \infty$. Therefore in the hard wall
approximation an effect of the second contribution $V_{SO}^{\gamma}$
limits to zero.

Next, for numerical computation of the transmission through the
semiconductor heterostructure we assume a connection at least to
two electrodes in which there is no the SOI. Then we can specify
the electron state by quantum numbers, the number of the energy
subband $n$ and spin projection $\sigma=\sigma_z$. This assumption
implies that far from waveguides or quantum dots the SOI constant
$\alpha$ equals zero in the electrodes. Neglecting by real space
behavior of the microscopic electric field at the edge of the
heterestructure we assume that the field is directed normal to the
plane of the heterostructure everywhere and has a stepwise
behavior at the edges. As a result we obtain the stepwise behavior
for the Razhba SOI constant $\alpha$. Such a model was used by Hu
and Matsuyama \cite{hu}. Similar to (\ref{soi2dd}) we obtain that
the third contribution to the SOI takes the following form
\begin{equation}
\label{SOI2d} V_{SO}^{\alpha\alpha} =\,-\hbar^2
\frac{i}{2}\left(\sigma_y \frac{\partial{\alpha}}{\partial
x}-\sigma_x \frac{\partial{\alpha}}{\partial y}\right).
\end{equation}

\section{The transmission through billiard with the SOI}

In this section we prove that the SOI gives no spin polarization
for electron transmission through arbitrary billiards if energy of
incident electron belongs to the first energy subband. In
dimensionless form the stationary Schr\"{o}dinger equation has the
following form
\begin{equation}
\label{SE} \left\{-\nabla^2 +
v_{SO}\right\}\psi=\epsilon\psi,\quad \psi=
\left(\matrix{u_1(x,y)\cr u_2(x,y)\cr}\right).
\end{equation}
Here $\epsilon=E/E_0, E_0=\frac{\hbar^2}{2m^{*}L^2}$,
\begin{equation}
\label{soi2d}
 v_{SO} =\beta\left(i\sigma_x\frac{\partial}{\partial y}-
 i\sigma_y\frac{\partial}{\partial x} \right) -
\frac{i}{2}\left(\sigma_y \frac{\partial{\beta}}{\partial
x}-\sigma_x \frac{\partial{\beta}}{\partial y}\right),
\end{equation}
$L$ is a characteristic scale of the system, $\beta=2m^{*}\alpha
L$  is the dimensionless SOI constant. Correspondingly in Eqs
(\ref{SE}) and (\ref{soi2d}) coordinates $x, y$ are also
dimensionless.

\begin{figure}[t]
\includegraphics[width=.4\textwidth]{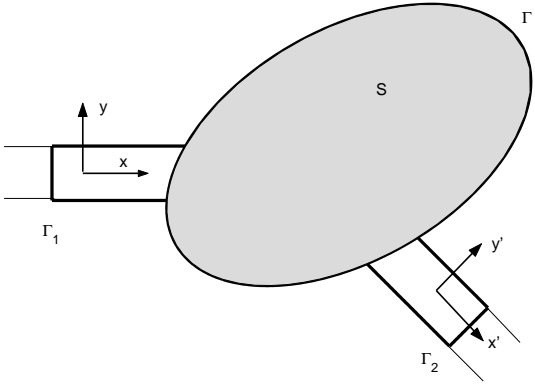}
\caption{Schematical view of two-dimensional billiard with two
attached leads. Dashed area shows a region with the SOI. The area
$S$ has boundary $\Gamma$ which crosses input lead and output one
at $\Gamma_1$ and $\Gamma_2$ respectively.}
\label{curvfig1}
\end{figure}

Let $S$ be an area of the structure under consideration which
involves a billiard the SOI and leads as shown in Fig.
\ref{curvfig1}.

Let $\Gamma$ denote a boundary which crosses input
lead and output one at $\Gamma_1$ and $\Gamma_2$ respectively. We
suppose that there is no spin-orbit interaction in the leads, {\it
i.e} $\beta=0$ at $\Gamma_i, i=1,2$. At the rest of boundary
$\Gamma$ we imply the Dirichlet boundary conditions for solution
of Schr\"{o}dinger equation (\ref{SE}) $\psi\vert_{_{\Gamma}}=0$.
As the scale $L$ we take $L=d$.

Therefore we can write solution in the electrodes as follows
\begin{eqnarray}
\label{waves} |inc,n,\sigma>=\sqrt{2}\sin(\pi n y)\exp(ik_n
x)|\sigma>,\nonumber\\ |refl,n,\sigma>=\sqrt{2}\sum_{m,\sigma'}
r_{mn,\sigma,\sigma'}\sin(\pi m y)\exp(-ik_m
x)|\sigma'>,\nonumber\\ |tr,n,\sigma>=\sqrt{2}\sum_{m,\sigma'}
t_{mn,\sigma,\sigma'}\sin(\pi m y')\exp(ik_m
x')|\sigma'>\nonumber\\
\end{eqnarray}
where $|\sigma>$ is the spin states defined spin projection along
some axis, say, the z-axis. The energy is
\begin{equation}
\label{kn} \epsilon=k_n^2+\pi^2 n^2
\end{equation}
where $n=1, 2, \ldots $ numerates a number of the energy subbands.

Introduce complex derivatives
\begin{equation}
\label{zcoord} \frac{\partial}{\partial z} =
\frac{1}{2}\left(\frac{\partial}{\partial
x}-i\frac{\partial}{\partial y}\right)
\end{equation}
we write the Schr\"{o}dinger equation (\ref{SE}) as follows
\begin{eqnarray}
\label{u1u2} \left(\frac{\partial^2}{\partial z \partial z^{*}}+
\frac{1}{4}\epsilon\right)u_1 + \frac{\beta}{2}
\frac{\partial{u_2}}{\partial z} + \frac{u_2}{4}
\frac{\partial{\beta}}{\partial z}=0,\nonumber\\
\left(\frac{\partial^2}{\partial z \partial
z^{*}}+\frac{1}{4}\epsilon\right)u_2 - \frac{\beta}{2}
\frac{\partial{u_1}}{\partial {z^{*}}} - \frac{u_1}{4}
\frac{\partial{\beta}}{\partial {z^{*}}}=0,
\end{eqnarray}
where $u_1, u_2$ are the components of the spin state. Assume that
there is auxiliary degenerated state with components $v_1, v_2$.
In particular, it might be the Kramers degenerated state. Then,
for these two states the Green formula follows
\begin{equation}
\label{green} \int_S (u\Delta v - v\Delta u)dS=
\oint_{\Gamma}\left(u \frac{\partial v}{\partial n} - v
\frac{\partial u}{\partial n}\right)dl
\end{equation}
where $n$ is an exterior normal to the boundary $\Gamma$.

From the Schr\"{o}dinger equation we have
\begin{eqnarray}
\label{uv} v_2\left(\frac{\partial^2}{\partial z \partial{z^{*}}}+
\frac{1}{4}\epsilon\right)u_1 + \frac{\beta}{2}
v_2\frac{\partial{u_2}}{\partial z} + \frac{1}{4} u_2 v_2
\frac{\partial{\beta}}{\partial z}=0,\nonumber\\
u_2\left(\frac{\partial^2}{\partial z \partial{z^{*}}}+
\frac{1}{4}\epsilon\right)v_1 + \frac{\beta}{2}
u_2\frac{\partial{v_2}}{\partial z} + \frac{1}{4} u_2 v_2
\frac{\partial{\beta}}{\partial z}=0,\nonumber\\
u_1\left(\frac{\partial^2}{\partial z \partial{z^{*}}}+
\frac{1}{4}\epsilon\right)v_2 - \frac{\beta}{2}
u_1\frac{\partial{v_1}}{\partial z^{*}} - \frac{1}{4} u_1 v_1
\frac{\partial{\beta}}{\partial z^{*}}=0,\nonumber\\
v_1\left(\frac{\partial^2}{\partial z \partial{z^{*}}}+
\frac{1}{4}\epsilon\right)u_2 - \frac{\beta}{2}
v_1\frac{\partial{u_1}}{\partial z^{*}} - \frac{1}{4} u_1 v_1
\frac{\partial{\beta}}{\partial z^{*}}=0.
\end{eqnarray}
Combining each couple of equations in (\ref{uv}) we obtain
$$ v_2\frac{\partial^2{u_1}}{\partial z \partial{z^{*}}}
+u_2\frac{\partial^2{v_1}}{\partial z \partial{z^{*}}}+
\frac{1}{4}\epsilon (u_1 v_2+u_2 v_1)+
\frac{1}{2}\frac{\partial{(\beta u_2 v_2)}}{\partial z} =
0,$$
\begin{equation}
\label{uv1}
 u_1\frac{\partial^2{v_2}}{\partial z
\partial{z^{*}}}+v_1\frac{\partial^2{u_2}}{\partial z
\partial{z^{*}}}+\frac{1}{4}\epsilon (u_1 v_2+u_2 v_1)
-\frac{1}{2}\frac{\partial{(\beta u_1
v_1)}}{\partial{z^{*}}}=0.\nonumber\\
\end{equation}
Extracting the second equation from the first one in (\ref{uv1})
we obtain
$$ v_2\frac{\partial^2{u_1}}{\partial z \partial z^{*}} -
u_1\frac{\partial^2{v_2}}{\partial z \partial z^{*}} +
u_2\frac{\partial^2{v_1}}{\partial z \partial z^{*}} -
v_1\frac{\partial^2{u_2}}{\partial z \partial z^{*}} $$
\begin{equation}
\label{uv2}
+ \frac{1}{2} \frac{\partial{(\beta u_1
v_1)}}{\partial z^{*}} + \frac{1}{2}\frac{\partial{(\beta u_2
v_2)}}{\partial z}=0.
\end{equation}

Integration of this equation over the billiard area $S$ with use
of the Green formula (\ref{green}) gives the following $$
\oint_{\Gamma}\left(v_2\frac{\partial{u_1}}{\partial
n}-u_1\frac{\partial{v_2}}{\partial n}   \right)\,dl+
 \oint_{\Gamma}\left(u_2\frac{\partial{v_1}}{\partial n}-v_1\frac{\partial{u_2}}{\partial n}
 \right)\,dl
 $$
\begin{equation}
\label{green1} +\int_{S}\frac{\partial{(2 \beta u_1
v_1)}}{\partial z^{*}}
 \,dS+\int_{S}\frac{\partial{(2 \beta u_2 v_2)}}{\partial z}\,dS=0.
\end{equation}
Since at $\Gamma$ either $u_1 = 0, v_1 = 0$ , or $\beta = 0$, the
last two integrals in (\ref{green1}) equal zero and formula
(\ref{green1}) can be rewritten as follows
$$ \sum_{i=1,2}
\int_{\Gamma_i}\left(v_2\frac{\partial{u_1}}{\partial
n}-u_1\frac{\partial{v_2}}{\partial n}   \right)\,dl $$
\begin{equation}
\label{green2} + \sum_{i=1,2}
\int_{\Gamma_i}\left(u_2\frac{\partial{v_1}}{\partial
n}-v_1\frac{\partial{u_2}}{\partial n}
 \right)\,dl = 0.
\end{equation}
This formula is  sufficient to establish some symmetry rules
between ingoing and outgoing states. Let us consider the
first-channel transmission for $\epsilon < 4\pi^2$. In order to
ignore evanescent modes we will consider that boundaries
$\Gamma_i$ cross the leads far from the scattering region as shown
in Fig. \ref{curvfig1} . Let electron incidents from the input
lead being completely spin polarized up. It means that for the
incident state (\ref{waves}) $|\sigma>= \left(\matrix{1\cr
0\cr}\right)$. We denote the corresponding state interior the
structure $S$ as $\left(\matrix{u_{1\uparrow}(x,y)\cr
u_{2\uparrow}(x,y)\cr}\right)$ which is used as the u-solution in
Eq. (\ref{green2}). Correspondingly
$\left(\matrix{u_{1\downarrow}(x,y)\cr
u_{2\downarrow}(x,y)\cr}\right)$ denotes the v-solution in Eq.
(\ref{green2}) for the case of electron incidenting with spin
polarized down . We suppose that the boundaries $\Gamma_1$ and
$\Gamma_2$ cross the leads normally the leads and the x-axis is
parallel to the leads. Hence the normal $n$ is parallel to the
x-axis. Then from (\ref{waves}) at the boundary $\Gamma_2$ which
crosses the output lead we obtain the following relations
\begin{equation}
\label{Gamma2} \frac{\partial f }{\partial n}=ik_1 f
\end{equation}
where function $f$ refers to all components $u_{1\uparrow}, u_{2\uparrow},
u_{1\downarrow}, u_{2\downarrow}$.

These relations allow to exclude the boundary $\Gamma_2$ from
(\ref{green2}). At the boundary $\Gamma_1$ which crosses the
input lead we have
\begin{eqnarray}
\label{Gamma1} \frac{\partial{u_{1\uparrow}}}{\partial
n}=ik_1u_{1\uparrow} -2ik_1\sin(\pi y),\nonumber\\
\frac{\partial{u_{2\uparrow}}}{\partial
n}=ik_1u_{2\uparrow},\nonumber \\
\frac{\partial{u_{1\downarrow}}}{\partial n}=ik_1u_{1\downarrow},
\nonumber\\ \frac{\partial{u_{2\downarrow}}}{\partial
n}=ik_1u_{2\downarrow}-2ik_1\sin(\pi y).
\end{eqnarray}
We imply here that the origin of the $x, y$ coordinate system is at the
boundary $\Gamma_1$. Substituting the relations (\ref{Gamma1})
into Eq (\ref{green2}) we obtain
\begin{equation}
\label{Gamma1next}
\int_{\Gamma_1}\left(u_{1\uparrow}-u_{2\downarrow} \right)
\sin(\pi y) \,dy=0.
\end{equation}

Since at the boundary $\Gamma_1$ $$
u_{1\uparrow}=\tilde{u}_{1\uparrow} (x) \sin( \pi y), \\
u_{2\downarrow}=\tilde{u}_{2\downarrow} (x) \sin( \pi y). $$ we
obtain from (\ref{Gamma1next}) $$ u_{1\uparrow}
=u_{2\downarrow}.$$
Thus from (\ref{waves}) it follows that
amplitudes of the reflection
\begin{equation}
\label{sym1} r_{\uparrow,\uparrow}=r_{\downarrow,\downarrow}.
\end{equation}

Next, we take that the state $\left(\matrix{u_1\cr
u_2\cr}\right)=\left(\matrix{u_{1\uparrow}\cr
u_{2\uparrow}\cr}\right)$  coincides with the
state $\left(\matrix{v_1\cr
v_2\cr}\right)$ in (\ref{green2}). Then equation (\ref{green2}) simplifies as
follows $$
\sum_{i=1,2}\int_{\Gamma_i}\left(u_2\frac{\partial{u_1}}{\partial
n}-u_1\frac{\partial{u_2}}{\partial n}
 \right)\,dl = 0.$$
Substituting into this formula relations (\ref{Gamma1}) we obtain
\begin{equation}
-2 i k\int_{\Gamma_1}u_{2\uparrow} \sin(\pi y) \,dy=0.
\end{equation}
It gives us that $u_{2\uparrow} = 0$ or according to (\ref{waves})
$r_{\uparrow\downarrow}=0$ . Also similarly we obtain that
$u_{1\downarrow}=0$ at the boundary $\Gamma_1$. Thus we can write
the second symmetry rule for reflection amplitudes
\begin{equation}
\label{sym2} r_{\uparrow,\downarrow} = r_{\downarrow,\uparrow} =
0.
\end{equation}
From symmetry rules (\ref{sym1}) and (\ref{sym2}) and from the
current preservation it follows that the transmission
probabilities
\begin{equation}
\label{sym3} T_{\sigma}=\sum_{\sigma'}|t_{\sigma,\sigma'}|^2=T
\end{equation}
do not depend on the spin polarization of incident electron.

Till now we considered incident waves as spin polarized along the
z-axis at the boundary $\Gamma_1$. Let now consider a flow of
incident electrons which have no averaged spin polarization. In
particular we can present that half of electrons have the incident
state with spin up and half of electrons have the incident state
with spin down. Let us consider corresponding transmitted waves at
the boundary $\Gamma_2$. We prove that for a transmission through
the billiard with two attached leads there is no averaged spin
polarization, {\it i.e.} $<\sigma_{\alpha}> = 0,\, \alpha = x, y,
z$ if electron incidents being spin unpolarized in the first
energy subband. As previously we take the incident state in the
form (\ref{waves}) and write the states in leads as
\begin{equation}
\label{states} |\psi_{\uparrow}>=\left(\matrix{u_{1\uparrow}\cr
u_{2\uparrow}\cr}\right),\quad
|\psi_{\downarrow}>=\left(\matrix{u_{1\downarrow}\cr
u_{2\downarrow}\cr}\right)
\end{equation}
where the arrows up and down indicate that electron incident with
spins up and down. We take in the Green formulas (\ref{green2}) the first
function $u$ as $|\psi_{\uparrow}>$ and the second function $v$ as
$\hat{\sigma}_y\hat{C}|\psi_{\downarrow}>$ where $\hat{C}$ means a
complex conjugation. It means that the second function is
the Kramers degenerated state. Hence
\begin{equation}
\label{u1v1} \left(\matrix{u_1\cr
u_2\cr}\right)=\left(\matrix{u_{1\uparrow}\cr
u_{2\uparrow}\cr}\right),\quad \left(\matrix{v_1\cr
v_2\cr}\right)=\left(\matrix{iu_{2\downarrow}^{*}\cr
-iu_{1\downarrow}^{*}\cr}\right).
\end{equation}

Let us calculate integral (\ref{green2}). At a boundary $\Gamma_1$
crossing the input lead the integral equals zero since in the
input lead $u_2=0, v_1=0$. The second contribution into integral
(\ref{green2}) relates to the boundary $\Gamma_2$ crossing the
output lead. Using transmitted solution (\ref{waves}) one can
write Eq. (\ref{green2}) as follows
 $$
\int_{\Gamma_2}\left(v_2\frac{\partial{u_1}}{\partial
n}-u_1\frac{\partial{v_2}}{\partial n}   \right)\,dl+
 \int_{\Gamma_2}\left(u_2\frac{\partial{v_1}}{\partial n}-
 v_1\frac{\partial{u_2}}{\partial n}\right)\,dl
 $$
 $$ =
2ik\oint_{\Gamma_2}(v_2u_1-u_2v_1)dy = 0. $$ Therefore
$u_1v_2=u_2v_1$, or in terms of notations (\ref{u1v1})
\begin{equation}
\label{u2v2}
u_{1\uparrow}u_{1\downarrow}^{*}=-u_{2\uparrow}u_{2\downarrow}^{*}.
\end{equation}
From (\ref{u2v2}) it obviously follows $$
|u_{1\uparrow}||u_{1\downarrow}|=|u_{2\uparrow}||u_{2\downarrow}|.
$$ Moreover relation (\ref{sym3}) implies that $$
|u_{1\uparrow}|^2+|u_{2\uparrow}|^2=|u_{1\downarrow}|^2+
|u_{2\downarrow}|^2 $$ From these two relations one can obtain
that
\begin{equation}
\label{moduv} |u_{2\uparrow}|=|u_{1\downarrow}|,
|u_{1\uparrow}|=|u_{2\downarrow}|.
\end{equation}
Finally relations (\ref{u2v2}) and (\ref{moduv}) give
\begin{equation}
\label{u3v3}
u_{1\uparrow}u_{2\uparrow}^{*}=-u_{1\downarrow}u_{2\downarrow}^{*}.
\end{equation}

Mean values of spin components in corresponding states
(\ref{states}) are the following
\begin{eqnarray}
\label{sigmaxyz}
<\sigma_x>_{\uparrow}=Re(u_{1\uparrow}u_{2\uparrow}^{*}),\,
<\sigma_y>_{\uparrow}=Im(u_{1\uparrow}u_{2\uparrow}^{*}),\nonumber\\
<\sigma_z>_{\uparrow}=|u_{1\uparrow}|^2-|u_{2\uparrow}|^2,\nonumber\\
<\sigma_x>_{\downarrow}=Re(u_{1\downarrow}u_{2\downarrow}^{*}),\,
<\sigma_y>_{\downarrow}=Im(u_{1\downarrow}u_{2\downarrow}^{*}),
\nonumber\\
<\sigma_z>_{\downarrow}=|u_{1\downarrow}|^2-|u_{2\downarrow}|^.
\end{eqnarray}
Eqs (\ref{moduv}) - (\ref{sigmaxyz}) give rise to
\begin{equation}
\label{bfS} <\sigma_{\alpha}>_{\uparrow} =
-<\sigma_{\alpha}>_{\downarrow},\, \alpha = x, y, z,
\end{equation}
{\it i.e.} the spin polarizations are exactly opposite in sign for
transmission of electrons incidenting in corresponding spin
polarized states.

\begin{figure}[t]
\includegraphics[width=.4\textwidth]{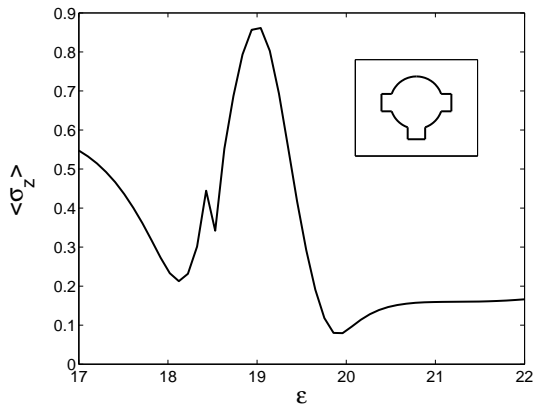}
\caption{Spin polarization of electrons transmitted through the
three terminal quantum dot versus energy of electron in the first
energy subband. An inset above shows a geometry of the structure.}
\label{curvfig2}
\end{figure}

Thus, for the transmission through any billiard with the SOI with
two attached leads the spin polarization does not exist if
the flow of electrons incidents in the first energy subband and have
no spin polarization.
Also, if there is no intersubband transmissions $t_{mn,\sigma
\sigma'}=0, m\neq n$, the spin polarization equals zero for
arbitrary energy. It takes place approximately, for example, for
adiabatic structures similar to curved waveguides (section V).
However in a vicinity of edges of the energy subbands $\pi^2 n^2$
the SOI gives rise to intersubband mixing. As a result we obtain
in numerical calculations strong spin polarization near the edges.
Moreover, if the billiard is connected to three or more leads, the
spin polarization of transmitted electrons exists even for the
transmission in the first energy subband. The effect of the third
lead  is demonstrated in Fig. \ref{curvfig2}.
Hence this effect
propose a way of the spin transistor complimentary to the way
proposed by Datta and Das \cite{datta}. The spin polarization of
transmitted electrons can be governed by a value of connection of
the third lead with the quantum dot. The most simple way is to
apply local electric field in the vicinity of the connection which
implies potential barrier closing the connection of the dot with
the third lead.


\section{The one dimensional curved wire}

\begin{figure}[t]
\includegraphics[width=.4\textwidth]{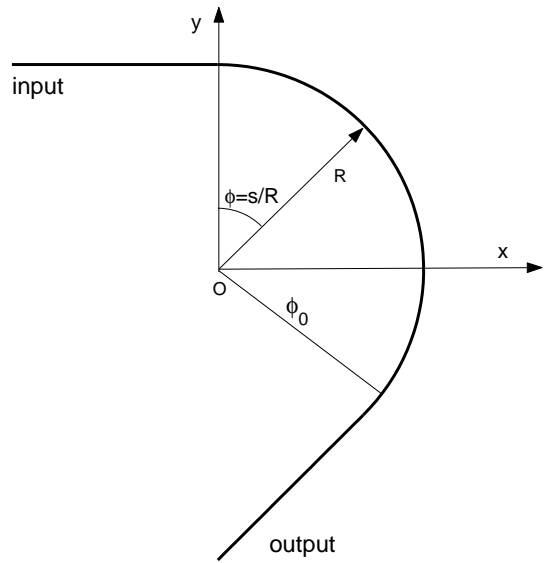}
\caption{Schematical view of one-dimensional curved wire.}
\label{curvfig3}
\end{figure}

A model in which only the single channel transmission takes place
is the one-dimensional wire. Therefore for a transmission through
the one-dimensional wire of any form the SOI can not give rise to
the spin polarization. However this model is interesting by that
allows to find spin evolution analytically. A case of straight
wire was considered by \cite{datta,mireles}. Here we consider a
curved wire consisted of a segment of circle with radius $R$
attached to infinite straight one dimensional wires as shown in
Fig. \ref{curvfig3}.

\begin{figure}[t]
\includegraphics[width=.4\textwidth]{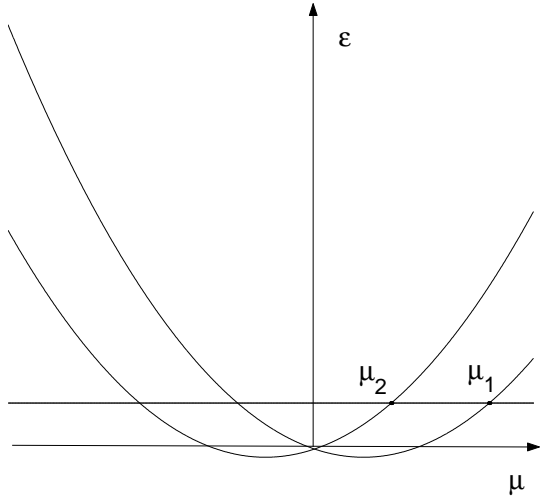}
\caption{The energy spectrum defined by formula (\ref{energy}) for
$\alpha = 1$ . Values of $\mu$ corresponded to clockwise movement
of electron along the curved wire are shown by thick points.}
\label{curvfig4}
\end{figure}

We take a length of the segment as $L=\phi_0
R$ and a position coordinate as $s=\phi R$. The Hamiltonian of the
wire has the following form \cite{qian,chaplik} $$
H=\frac{\hbar^2}{2m^{*}R^2}\widetilde{H},$$
\begin{equation}
\label{Hseg} \widetilde{H}=\left[\frac{\partial}{i\partial{\phi}}
+\frac{\beta}{2}\left(\sigma_y\cos\phi+
\sigma_x\sin\phi\right)\right]^2-\frac{\beta^2}{4},
\end{equation}
where $\beta=2m^{*}\alpha R$ is the dimensionless SOI constant.
Since $[J_z, H]=0$ where
$J_z=-i\frac{\partial}{\partial{\phi}}-\frac{1}{2}\sigma_z$, a
particular solution of the stationary Shcr\"odinger equation
$\widetilde{H}|\psi>=\epsilon|\psi>$ has the following form
\cite{qian,chaplik,magarill}
\begin{equation}
\label{psi0} |\psi>=\left(\matrix{Ae^{i\mu\phi}\cr Be^{i(\mu
-1)\phi}\cr}\right).
\end{equation}
The parameter $\mu$ defines the dimensionless wave number as
$k=\mu/R$ and is arbitrary until the boundary conditions are
imposed. Substituting the state (\ref{psi0}) into the
Shcr\"odinger equation one can obtain the following relation
between the energy of electron $\epsilon$ and the wave number
$\mu$
\begin{equation}
\label{ener}
(\epsilon-\mu^2)(\epsilon-(\mu-1)^2)-\beta^2(\mu-1/2)^2=0
\end{equation}
which gives
\begin{equation}
\label{energy} \epsilon_{\nu}=(\mu-1/2)^2+1/4
+\nu|\mu-1/2|\sqrt{\beta^2+1/4},\quad \nu=\pm 1.
\end{equation}
The spectrum (\ref{energy}) is shown in Fig. \ref{curvfig4}.

For fixed
energy $\epsilon$ Eq.(\ref{energy}) gives four solutions for the
wave number $\mu$. It is well known that \cite{davydov} for
electron transmission through potential profile a reflection is
negligibly small if the characteristic length of inhomogeneity much exceeds
the wave length (adiabatic regime). For our case we assume that
the radius of curvature of the wire is much larger in comparison
with the electron wave length. So we can ignore the reflection for
electron transmission through the quasi one-dimensional waveguide.

Since there is no reflection for transmission through the
one-dimensional waveguide we need only those values of the wave
number $\mu$ which correspond to clockwise movement of electron in
the waveguide. We denote its as $\mu_1, \mu_2$ shown in Fig.
\ref{curvfig4} . In what follows we use the following relation
between $\mu_1$ and $\mu_2$:
\begin{equation}
\label{deltamu} 2\lambda=\mu_2-\mu_1=-\sqrt{1+\beta^2}.
\end{equation}

Therefore general solution of the Shcr\"odinger equation for the
electron transmission without reflection can be written as follows
\begin{equation}
\label{psi} |\psi(\phi)>=\sum_{\nu=1,2} a_{\nu}
e^{i(\mu_{\nu}-1/2)\phi}U(\phi)\left(\matrix{A_{\nu}\cr
B_{\nu}\cr}\right)
\end{equation}
where
\begin{equation}
\label{A1}
 \left(\begin{array}{cc} 1-\sqrt{1+\beta^2} &
-i\beta\\ i\beta & -1-\sqrt{1+\beta^2} \\
\end{array}\right)\left(\matrix{A_2\cr B_2\cr}\right)=0,
\end{equation}
\begin{equation}
\label{A2}
 \left(\begin{array}{cc} 1+\sqrt{1+\beta^2} &
-i\beta\\ i\beta & -1+\sqrt{1+\beta^2} \\
\end{array}\right)\left(\matrix{A_1\cr B_1\cr}\right)=0.
\end{equation}

The matrix $U(\phi)$ has the following form
\begin{equation}
\label{U}
U(\phi)\, = \, \left(\begin{array}{cc} \exp(i\phi/2) &
0\\0 & \exp(-i\phi/2) \\
\end{array}\right).
\end{equation}

Evolution of the electron state (\ref{psi}) as length $s=\phi R$
of the curved wire can be presented as
$$|\psi(\phi)>=e^{i(\mu_1+\mu_2-1)\phi/2}U(\phi)\Lambda
$$
\begin{equation}
\label{psi1} \left(\begin{array}{cc} \exp(i\lambda \phi) & 0\\0 &
\exp(-i\lambda \phi)
\\
\end{array}\right)\Lambda^{-1}|\psi(0)>
\end{equation}
where
\begin{equation}
\label{Lambda} \Lambda\, = \, \left(\begin{array}{cc} A_2 &
A_1\\B_2 & B_1 \\
\end{array}\right).
\end{equation}

 From Eqs (\ref{A1}) and (\ref{A2}) we can rewrite (\ref{Lambda}) as
follows
\begin{equation}
\label{Lambda1} \Lambda\, = \, \left(\begin{array}{cc} A_2 &
-B^{*}_2\\B_2 & A^{*}_2 \\
\end{array}\right)
\end{equation}
and
\begin{equation}
\label{AB1} \frac{A_2}{B_2}=\frac{i\beta}{(1-\sqrt{1+\beta^2})}.
\end{equation}

Eq. (\ref{psi1}) can be presented as
$|\psi(\phi)>=T(\phi)|\psi(0)>$ which shows that the unitary
matrix $T(\phi)$ has a meaning of the transfer one. Since the
state $|\psi(\phi)>$ is spinor one the matrix $T(\phi)$
corresponds to rotation matrix for transport of electron along the
wire. In general case the rotation matrix is given by the Euler
angles ($\varphi, \theta, \gamma$) \cite{messia}
\begin{equation}
R(\varphi, \theta, \gamma)\, =
\,e^{-i\varphi\frac{1}{2}\sigma_z}e^{-i\theta\frac{1}{2}\sigma_y}
e^{-i\gamma\frac{1}{2}\sigma_z}.
\end{equation}
The rotation matrix $R$ has the following form
$$ R(\varphi, \theta, \gamma)$$
\begin{equation}
\label{rotate}
 =\left(\begin{array}{cc}
\exp(-i\frac{1}{2}(\varphi+\gamma))\cos(\frac{1}{2}\theta) &
-\exp(-\frac{1}{2}i(\varphi-\gamma))\sin(\frac{1}{2}\theta)\\
\exp(i\frac{1}{2}(\varphi-\gamma))\sin(\frac{1}{2}\theta) &
\exp(i\frac{1}{2}(\varphi+\gamma))\cos(\frac{1}{2}\theta)\\
\end{array}\right).
\end{equation}
In order to find the Euler angles let us consider to which
rotation corresponds  matrix
\begin{equation}
\label{aux} \Lambda\left(\begin{array}{cc} \exp(i\lambda \phi) &
0\\0 & \exp(-i\lambda \phi)
\\
\end{array}\right)\Lambda^{-1}.
\end{equation}
If the matrix $\Lambda$ were unit, the matrix
$$\left(\begin{array}{cc} \exp(i\lambda \phi) & 0\\0 &
\exp(-i\lambda \phi)\\
\end{array}\right)$$
would correspond to the rotation by the angle $\theta=-2\lambda
\phi$ around the z-axis. The matrix $\Lambda$ in (\ref{aux}) gives
rise to the clockwise rotation around the x-axis by the angle
$\theta$ which satisfies   to the following equation
\begin{equation}
\label{angle} \frac{A_2}{B_2}=\cot(\theta/2) e^{-i\varphi}=
\frac{i\beta}{1-\sqrt{1+\beta^2}}.
\end{equation}
In order to fulfill this equation we choose
\begin{equation}
\label{angles} \varphi=\pi/2,\quad \cot\frac{\theta}{2}=
\frac{\beta}{\sqrt{1+\beta^2}-1}.
\end{equation}
The angle $\gamma$ is remaining undefined. Below we put
$\gamma=0$. Let us choose new axis $z'$ in the $(y, z)$ plane with
the angle $\theta$ as the angle between the z-axis and $z'$-axis.
Thus, the full rotation matrix consists of the anticlockwise
rotation around the the $z'$-axis by the angle
$\phi\sqrt{1+\beta^2}$ and the clockwise rotation by the angle
$\phi$ around the z-axis. The last statement follows from the
matrix $U(\phi)$ in Eq.(\ref{psi1}). A knowledge of the evolution
of the spin state (\ref{psi1}) as dependent on the length $\phi R$
of the curved waveguide gives us a possibility to calculate in
particular evolution of the spin components. The result of
calculation is shown in Fig. \ref{curvfig6} by squares, triangles
and circles for spin components $\sigma_z, \sigma_x, \sigma_y$
respectively. As seen the results of numerical computation for the
two-dimensional curved waveguide are surprisingly close to present
one-dimensional model.


 \section{The two-dimensional curved waveguide}

\begin{figure}[t]
\includegraphics[width=.4\textwidth]{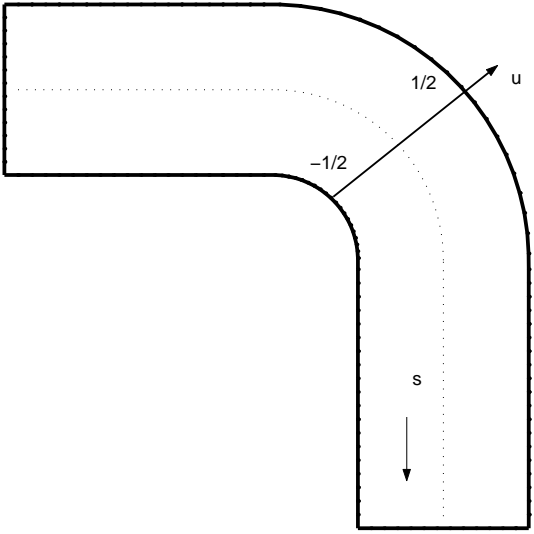}
\caption{A fragment of two-dimensional curved wire with width $d =
1$.}
\label{curvfig5}
\end{figure}

For consideration of the two-dimensional curved waveguide we
introduce the curved coordinate system $(s, u)$
\cite{exner,simanek} where $s$ is the coordinate of central line
along of the waveguide shown in Fig. \ref{curvfig5}.

We express
the Hamiltonian of the waveguide in dimensionless form by
following way
$$H=\frac{\hbar^2}{2m^{*}d^2}(\widetilde{H}_0+v_{SO}) ,$$ where
\begin{equation}\label{laplace}
\widetilde{H}_0=-\Delta=-g^{-1/2}\frac{\partial}{\partial{s}}g^{-1/2}\frac{\partial}{\partial{s}}
-g^{-1/2}\frac{\partial}{\partial{u}}g^{1/2}\frac{\partial}{\partial{u}},
\end{equation}
and $d$ is a width of the waveguide.
In what follows we consider a segment of the two-dimensional ring
with constant curvature $\gamma = 1/R$ attached to straight leads with the
same width as shown in Fig. \ref{curvfig5}. Therefore for the
segment we can write
\begin{eqnarray}\label{xy}
  x=a(s)-ub'(s)\nonumber\\
y=b(s)+ua'(s)\nonumber\\
a(s)=-R\cos(s/R), b(s)=R\sin(s/R)\nonumber\\
g^{1/2}=1+u\gamma(s)=\frac{u+R}{R}
\end{eqnarray}
with $\gamma(s)$ as a curvature of the curved waveguide which is
taken below constant. The SOI takes the following form at the
curved part of the waveguide $$ v_{SL} = $$
\begin{equation}
\label{soiseg} \beta\left(\begin{array}{cc} 0 &
e^{is/R}\left(\frac{\partial}{\partial{u}}+
ig^{-1/2}\frac{\partial}{\partial{s}}\right)\\
-e^{-is/R}\left(\frac{\partial}{\partial{u}}-
ig^{-1/2}\frac{\partial}{\partial{s}}\right) & 0\\
\end{array}\right).
\end{equation}
At the leads we assume that there is no the spin-orbital
interaction ($\beta=0$) as well as $\gamma=0, g^{1/2}=1$.

The Shcr\"odinger equation $$
\tilde{H}\left(\matrix{\psi_{\uparrow}\cr
\psi_{\downarrow}\cr}\right)=\epsilon\left(\matrix{\psi_{\uparrow}\cr
\psi_{\downarrow}\cr}\right)$$ with the total Hamiltonian as
$\tilde{H}=\tilde{H}_0+v_{SO}$ takes the following form
$$ g^{-1/2}\frac{\partial}{\partial{s}}
\left(g^{-1/2}\frac{\partial{\psi_{\uparrow}}}{\partial{s}}\right)
+g^{-1/2}\frac{\partial}{\partial{u}}
\left(g^{1/2}\frac{\partial{\psi_{\uparrow}}}{\partial{u}}\right)
$$ $$ + \epsilon\psi_{\uparrow}-\beta
e^{is/R}\left(\frac{\partial{\psi_{\downarrow}}}{\partial{u}}+
ig^{-1/2}\frac{\partial{\psi_{\downarrow}}}{\partial{s}}\right)=0,$$
$$ g^{-1/2}\frac{\partial}{\partial{s}}
\left(g^{-1/2}\frac{\partial{\psi_{\downarrow}}}{\partial{s}}\right)
+g^{-1/2}\frac{\partial}{\partial{u}}
\left(g^{1/2}\frac{\partial{\psi_{\downarrow}}}{\partial{u}}\right)
$$
\begin{equation}
\label{down}
 + \epsilon\psi_{\downarrow}+\beta
e^{-is/R}\left(\frac{\partial{\psi_{\uparrow}}}{\partial{u}}-
ig^{-1/2}\frac{\partial{\psi_{\uparrow}}}{\partial{s}}\right)=0,
\end{equation}
The solutions of Eqs (\ref{down}) which satisfy to the Dirichlet
boundary conditions ($u=\pm 1/2$) can be presented as
\cite{exner,simanek}
\begin{eqnarray}\label{solution}
\psi_{\uparrow}(u,s)=\sum_{n=1}^{\infty}A_{\uparrow n}(s) \sin(\pi
n(u+1/2))\nonumber\\
\psi_{\downarrow}(u,s)=\sum_{n=1}^{\infty}A_{\downarrow n}(s)
\sin(\pi n(u+1/2)).
\end{eqnarray}
Substitution of (\ref{solution}) into Eqs
(\ref{down}) gives $$ \sum_{n=1}^{\infty}[L_{mn}A_{\uparrow
n}^{''}(s)+P_{mn}A_{\uparrow n}(s)-\beta e^{i\gamma
s}Q_{mn}A_{\downarrow n}(s)$$
$$- i\beta e^{i\gamma s}R_{mn}A_{\downarrow n}^{'}(s)]=[(\pi
m)^2-\epsilon]A_{\uparrow m} $$ $$
\sum_{n=1}^{\infty}[L_{mn}A_{\downarrow
n}^{''}(s)+P_{mn}A_{\downarrow n}(s)+\beta e^{-i\gamma
s}Q_{mn}A_{\uparrow n}(s) $$
\begin{equation}
\label{coef} - i\beta e^{-i\gamma s}R_{mn}A_{\uparrow
n}^{'}(s)]=[(\pi m)^2-\epsilon]A_{\downarrow m}.
\end{equation}
Here we introduced the following notations
\begin{eqnarray}\label{coef1}
L_{mn}=2\int_{-1/2}^{1/2} \frac{\sin(\pi m(u+1/2))\sin(\pi
n(u+1/2))}{(1+u\gamma)^2}du,\nonumber\\ P_{mn}=2\pi n\int_{-1/2}^{1/2}
\frac{\gamma\sin(\pi m(u+1/2))\cos(\pi
n(u+1/2))}{1+u\gamma}du,\nonumber\\ R_{mn}=2\int_{-1/2}^{1/2}
\frac{\sin(\pi m(u+1/2))\sin(\pi
n(u+1/2))}{1+u\gamma}du,\nonumber\\
 Q_{mn}=2\pi n\int_{-1/2}^{1/2} \sin(\pi
m(u+1/2))\cos(\pi n(u+1/2))du.\nonumber\\
\end{eqnarray}
\section{Numerical results}
\begin{figure}[t]
\includegraphics[width=.4\textwidth]{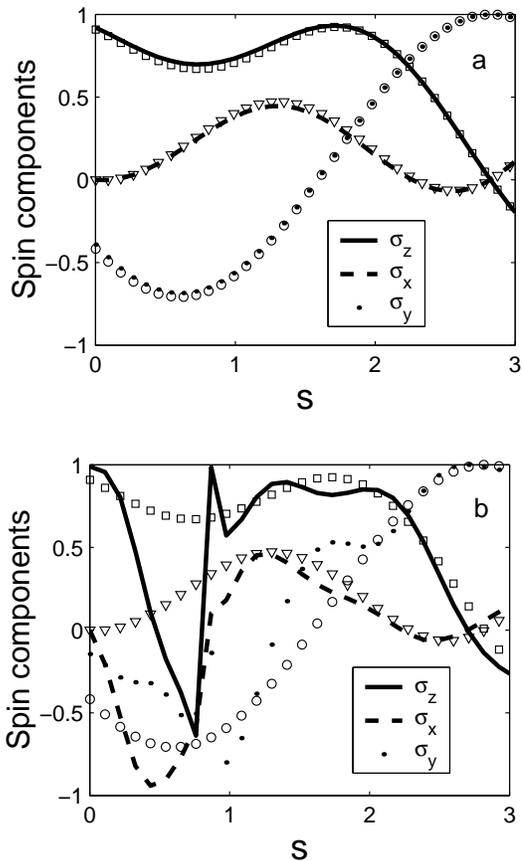}
\caption{The spin components as dependent on the length $s$ of a
curved two-dimensional waveguide. The result of calculation based
on the state (\ref{psi1}) for the curved one-dimensional wire is
shown by squares ($\sigma_z$), triangles ($\sigma_x$) and circles
($\sigma_y$). The radius of wire $R=d$ where $d$ is the width of
waveguide. The dimensionless spin-orbit constant
$\beta=2m^{*}\alpha d$ equals unit.(a) The dimensionless energy
$\epsilon=25$ (the first channel transmission) and (b)
$\epsilon=39.25$ (near an edge of the second subband).}
\label{curvfig6}
\end{figure}

In numerical practice we solve the system of Eqs (\ref{coef}) and
(\ref{coef1}) taking a finite number of waveguide modes. This
number of modes was controlled by the normalization condition that
sum of the total reflection probabilities and the total
transmission ones is to be equaled to unit. The spin components
$<\sigma_{\alpha}>$ were calculated at the attached outgoing
straight electrode in which we assumed there is no the spin-orbit
interaction by following formula
\begin{equation}
\label{avspin} <\sigma_{\beta}(s)>=\frac{\int_{-1/2}^{1/2} du
<\psi(u,s)|\hat{\sigma_{\beta}}|\psi(u.s)>}{\int_{-1/2}^{1/2} du
<\psi(u,s)|\psi(u.s)>}.
\end{equation}
In Fig.\ref{curvfig5} the outgoing electrode as well as incoming
one are not shown. Fig. \ref{curvfig6} shows evolution of the spin
components (\ref{avspin}) versus the longitudinal coordinate $s$.

\begin{figure}[t]
\includegraphics[width=.4\textwidth]{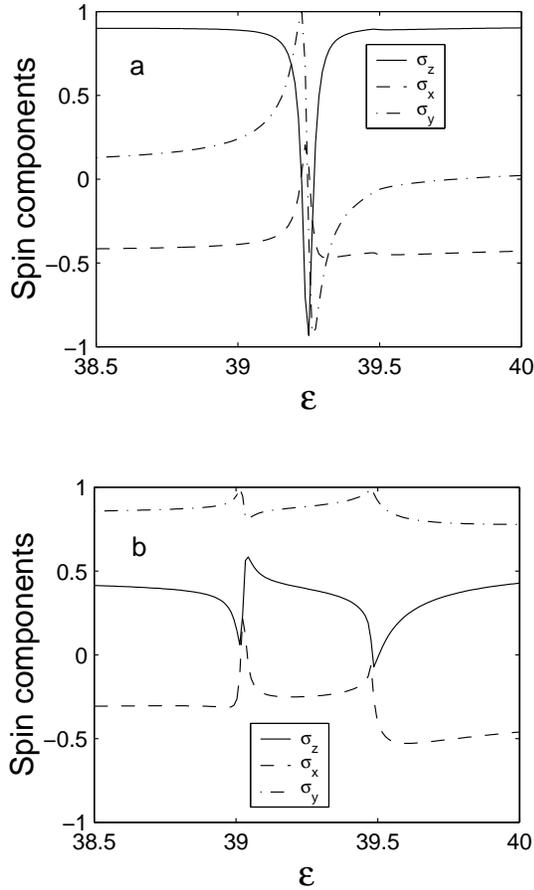}
\caption{The spin components as dependent on the energy of
incident electron for (a) $\phi_0 = 90^o$-curved waveguide and (b)
$\phi_0 = 180^o$ one. $\beta=1$.}
\label{curvfig7}
\end{figure}

It is surprising that for energy of incident electron far from the
edge of energy subband the spin evolution almost coincides with
the one-dimensional curved wire shown in Fig. \ref{curvfig6} by
squares, triangles and circles. In Fig. \ref{curvfig7} (a) the
energy dependence of the spin components are shown which
demonstrates remarkable phenomenon of spin flipping at the edge of
the second energy subband $E_2=(2 \pi)^2\approx 39.4$. It is
interesting that increasing of region with the SOI by increasing
of length curved waveguide or increasing of the spin-orbit
constant leads to double flipping of electron spin for
transmission through the waveguide as shown in Fig. \ref{curvfig7}
(b) . This phenomenon is a consequence of the intersubband mixing
by the SOI as it was discussed in section III.

Therefore one can expect strong deviation of the curved
two-dimensional waveguides from the one-dimensional one for the
spin evolution near edges of the subbands $\pi^2 n^2$. In fact,
one can see from Fig. \ref{curvfig7} (b) that for energy of
incident electron $E\approx 4\pi^2$ the spin evolution with length
of the curved waveguide is strongly deviates from the case of
one-dimensional curved wire.

\acknowledgments This work has been partially by RFBR Grant
01-02-16077 and the Royal Swedish Academy of Sciences.

$^{*}$ e-mails:almsa$@$ifm.liu.se, almas$@$tnp.krasn.ru

\end{document}